\documentclass[acmsmall]{acmart}

\usepackage{algorithm, algorithmic}
\usepackage{enumitem}
\usepackage[breakable]{tcolorbox}
\usepackage{braket}

\AtBeginDocument{%
  }

\setcopyright{none}

\begin{document}

\title{Structure-Aware Quantum Circuit Equivalence Checking via Sparse Pauli Propagation}


\author{Shangzhou Xia}
\authornote{Both authors contributed equally to this research.}
\orcid{0009-0006-2775-9633}
\affiliation{%
  \institution{Kyushu University}
  \state{Fukuoka}
  \country{Japan}
}
\email{xia.shangzhou.218@s.kyushu-u.ac.jp}

\author{Junjie Luo}
\authornotemark[1]
\orcid{0009-0008-7821-6879}
\affiliation{%
  \institution{RIKEN Center for Quantum Computing}
  \city{Wako}
  \state{Saitama}
  \country{Japan}
}
\email{junjie.luo@riken.jp}

\author{Qichen Wang}
\orcid{0009-0009-8268-7152}
\affiliation{%
  \institution{Kyushu University}
  \state{Fukuoka}
  \country{Japan}
}
\email{wang.qichen.256@s.kyushu-u.ac.jp}

\author{Yasunari Suzuki}
\orcid{0000-0002-8005-357X}
\affiliation{%
  \institution{RIKEN Center for Quantum Computing}
  \city{Wako}
  \state{Saitama}
  \country{Japan}}
\email{yasunari.suzuki@riken.jp}

\author{Jianjun Zhao}
\authornote{Corresponding author.}
\orcid{0000-0001-8083-4352}
\affiliation{%
  \institution{Kyushu University}
  \state{Fukuoka}
  \country{Japan}
}
\email{zhao@ait.kyushu-u.ac.jp}

\renewcommand{\shortauthors}{Shangzhou Xia, Junjie Luo, Qichen Wang, Yasunari Suzuki and Jianjun Zhao}

\begin{abstract}
Quantum circuit equivalence checking is essential for validating compiler optimizations, but its computational cost can vary substantially with circuit structure. We present a structure-aware equivalence checker for unitary quantum circuits based on sparse Pauli propagation. It combines two complementary verification strategies: generator propagation compares the conjugation actions of the circuits on Pauli generators, while Cut-Miter exploits cancellation between circuit products at selected cut points. A structure-aware verification procedure uses circuit characteristics to select between these strategies while explicitly accounting for numerical and approximation errors. We evaluate our framework against three existing equivalence checkers on two benchmark suites containing circuit pairs with up to 512 qubits. Across 434 scored circuit pairs, our framework produces reference-consistent decisions for 96.3\% of the pairs without any incorrect verdicts, while the evaluated baselines achieve at most 68.7\% reference-consistent decisions. It also achieves the lowest aggregate runtime among the evaluated tools, reducing total runtime by 48.5\%--76.0\% relative to these baselines. Ablation studies confirm the benefits of cut-based verification, Pauli-term pruning, and bounded propagation attempts. These results indicate that structure-aware sparse Pauli propagation, combined with complementary verification strategies and explicit error accounting, can provide an effective approach to practical quantum circuit equivalence checking.

\end{abstract}



\maketitle

\section{Introduction}
\label{sec:intro}
Quantum programs typically undergo rewriting and optimization during compilation~\cite{opt1, opt2}, including decomposing high-level operations, canceling or commuting gates, synthesizing alternative subcircuits, and translating circuits into hardware-supported gate sets. These transformations may substantially alter the syntax and structure of a circuit, but must preserve its semantics. However, several studies~\cite{qdiff, morphq,qemi,shakeup} have revealed semantic inconsistencies after such transformations in widely used quantum development frameworks. Consequently, semantic regression checking has become an important component of building reliable quantum software systems.

To address this problem, existing research has primarily pursued two approaches: testing and verification. Quantum transformation testing typically follows the metamorphic testing paradigm~\cite{morphq,qemi}. By executing the original and transformed circuits separately on quantum devices, one can obtain only a finite number of samples from their output distributions. However, matching the observed samples cannot establish equivalence for all possible inputs, and discrepancies may be masked by statistical noise arising from a finite number of shots as well as hardware errors, making it difficult to construct an effective test oracle. Quantum equivalence checkers typically avoid explicit state-vector simulation by leveraging representations such as decision diagrams, ZX diagrams, path sums, weighted model counting, and Pauli propagation~\cite{qcec1,qcec2,ecmc,quprs, thanos2023fast}. These techniques have enabled the verification of quantum circuits far beyond the scale tractable by dense-matrix comparison. Nevertheless, their performance remains highly dependent on circuit structure. A representation that is compact for one class of circuits may grow exponentially for another. More subtly, even when the final equivalence condition admits a compact representation, establishing that condition may require constructing intermediate states of enormous size. Consequently, relying on a single symbolic representation and a fixed propagation strategy can create a representation bottleneck: even when a sparse representation exists for the same verification problem, a checker may still fail if intermediate representations become dense, leading to memory exhaustion or excessive runtime.

Our key observation is that quantum-circuit equivalence exposes two complementary forms of sparsity. The first is \emph{action sparsity}. For Clifford-dominated circuits, the conjugation of a Pauli generator by the circuit remains compact. Equivalence can therefore be checked by comparing the images of the $2n$ generators without constructing either circuit's complete operator. The second is \emph{difference sparsity}. Circuits produced by compiler rewrites, or local mutations, frequently share substantial structure, even when each circuit is individually expensive to propagate. By constructing their miter $U_2^\dagger U_1$ from the two sides in an alternating order, common structure can cancel early, and the remaining difference may stay sparse. Neither form of sparsity dominates universally: Clifford circuits may have simple generator actions even when their Pauli expansions are exponentially large, whereas rotation-dense circuit pairs may admit compact difference operators even when independent propagation is expensive.

Based on this observation, we present a structure-aware equivalence checker for unitary quantum circuits. Instead of committing to one representation, our framework organizes these complementary Pauli-propagation engines as a bounded portfolio. 
\emph{Generator} backend independently propagates the Pauli generators through both circuits and compares their conjugation actions. 
\emph{Cut-Miter} backend searches several circuit cut points and compares partial products at a potentially sparser boundary and incrementally constructs a sparse Pauli representation of $U_2^\dagger U_1$, consuming operations from both circuits so that aligned structure is eliminated as early as possible. 
A lightweight router selects an initial engine using circuit characteristics, including the number and proportion of non-Clifford rotations. Failed fast paths consume only bounded slices of the available resources and fall back to more general verification paths. This portfolio is designed not only for performance but also for dependable use as a software-testing oracle. All engines decide the same semantic property using complete residual $L_2$ mass rather than individual Pauli coefficients, since a semantic difference may be distributed across many individually small coefficients. Our framework explicitly tracks the errors introduced by Pauli-term pruning and semantic lowering, and uses the resulting error bound to certify \textsc{Equivalent} or \textsc{Not-Equivalent} when the residual is sufficiently separated from the tolerance. Otherwise, it returns \textsc{Undecided} rather than treating numerical or resource limitations as evidence of equivalence. 

We evaluate our framework on 434 scored circuit pairs from two benchmark suites, with circuit widths ranging up to 512 qubits, against three existing equivalence checkers. Our framework returns correct, reference-consistent decisions for 96.3\% of the scored pairs with no incorrect verdicts, compared with at most 68.7\% correct decisions for the evaluated baselines. It also achieves the lowest aggregate runtime, reducing total runtime by 48.5\%--76.0\% relative to these tools. The results illustrate the benefit of adapting the verification strategy to circuit structure: cut-miter comparisons exploit cancellation between related circuits, whereas generator propagation can efficiently handle circuits with substantial Clifford structure. Ablation results further show that both cut-based verification and Pauli-term pruning are important for maintaining decision coverage and controlling computational cost, while bounded generator attempts prevent expensive propagation cases from consuming the entire verification budget. Finally, varying the pruning threshold from $10^{-12}$ to $10^{-14}$ changes the outcome of only one of 438 evaluated instances and introduces no EQ--NEQ switches, indicating that the observed decisions are stable across the tested range. These results support the effectiveness of a bounded, structure-aware portfolio for practical quantum circuit equivalence checking.

The contribution is listed as follows:

\begin{itemize}
    \item \noindent \textbf{A structure-aware equivalence checking method.}
    We combine Pauli-generator propagation with cut-point partial-product comparison to exploit compact circuit actions and cancellation between circuits. The method selects between these strategies according to circuit structure and resource limits.

    \item \noindent \textbf{Numerical decisions with approximation error bounds.}
    We formulate equivalence decisions using residual norms and accumulated errors from pruning and semantic lowering.

    \item \noindent \textbf{An implementation and empirical evaluation.}
    We implement our framework and evaluate it against three existing checkers on two benchmark suites, obtaining more correct decisions and lower aggregate runtime. Ablation and sensitivity studies examine the contributions of individual components and the effect of pruning thresholds.
\end{itemize}







The remainder of this paper is organized as follows. Section~\ref{sec:background} provides the necessary background. Section~\ref{sec:method} presents our methodology. Section~\ref{sec:experiment} describes the experimental setup and discusses the empirical results. Section~\ref{sec:validity} discusses the threats to validity, while Section~\ref{sec:related} reviews related work. Finally, Section~\ref{sec:conclusion} concludes the paper.

\section{Background}
\label{sec:background}

\subsection{Quantum Circuits and Unitary Semantics}
\label{subsec:circuit}

An $n$-qubit pure state is represented by a unit vector $\ket{\psi}\in\mathbb{C}^{2^n}$. A unitary gate acting on $k$ qubits is described by a $2^k\times 2^k$ unitary matrix and is embedded into the $n$-qubit state space according to the qubits on which it acts. Common single-qubit gates include the Pauli gates $X$, $Y$, and $Z$, the Hadamard gate $H$, and the rotation gates
\begin{equation}
    R_\sigma(\theta)=\exp(-i\theta\sigma/2),
\qquad \sigma\in\{X,Y,Z\}.
\end{equation}
Multi-qubit gates, such as the controlled-NOT gate $\mathrm{CX}$, introduce interactions between qubits.

We consider quantum circuits composed of unitary gates. If the gates of a circuit $C$ are applied in the order $G_1,G_2,\ldots,G_m$, its unitary semantics is $U_C=G_mG_{m-1}\cdots G_1$. Thus, for an input state $\ket{\psi}$, the circuit produces $U_C\ket{\psi}$. Because multiplication order is significant, two circuits may contain similar gates while implementing different unitary transformations.
A quantum state can also be described by a density operator $\rho$, a positive semidefinite matrix with trace one. For a pure state, $\rho=\ket{\psi}\bra{\psi}$. A circuit with unitary $U_C$ transforms it into $U_C\rho U_C^\dagger$.

An $n$-qubit Pauli string is a tensor product $P=P_1\otimes\cdots\otimes P_n$, where each $P_j$ belongs to $\{I,X,Y,Z\}$. We write $\mathcal{P}_n$ for the set of these strings. The strings $X_j$ and $Z_j$ act as $X$ and $Z$, respectively, on qubit $j$ and as $I$ on all other qubits. A Clifford gate maps each Pauli string under conjugation to another Pauli string, possibly with a sign change. Common Clifford gates include the Pauli gates, $H$, $S$, and $\mathrm{CX}$. Non-Clifford gates, including $T$ and generic rotations, can produce a linear combination of Pauli strings.

\subsection{Quantum Circuit Equivalence Checking}
\label{subsec:ec}

As quantum computing continues to advance, quantum circuits are increasingly processed by compilation and optimization techniques~\cite{opt3,opt4,opt5} to improve their execution efficiency and adapt them to the constraints of quantum hardware. These transformations may substantially change the structure of a circuit, producing circuit variants that are syntactically different from the original while being expected to implement the same quantum operation. Therefore, it is essential to verify that such transformations preserve the semantics of the original circuit. Quantum circuit equivalence checking~\cite{ecmc,qcec1,qcec2,quprs,ec1,ec2,ec3} addresses this need by determining whether two quantum circuits implement the same quantum operation, typically up to a global phase. It thus provides a fundamental correctness guarantee for quantum circuit compilation and optimization.

\begin{definition}[Equivalence]
\label{def:equiv}
Two $n$-qubit circuits $C_1$ and $C_2$ with unitaries $U_1$ and $U_2$ are equivalent, written $C_1\equiv C_2$, if $U_1=e^{i\varphi}U_2$ for some $\varphi\in[0,2\pi)$.
\end{definition}

The factor $e^{i\varphi}$ is a global phase and does not affect measurement probabilities. Since $U_2^\dagger U_2=I$, Definition~\ref{def:equiv} is equivalent to
\begin{equation}
    M=U_2^\dagger U_1=e^{i\varphi}I.
\end{equation}
The composite operator $M$ is called a \emph{miter}. It reduces equivalence checking to determining whether a single operator equals the identity up to a global phase.

\subsection{Pauli Propagation}
\label{subsec:pauli-choi}

Pauli propagation provides a compact way to represent and analyze quantum operators without explicitly constructing their exponentially large matrix representations. Let $\mathcal{P}_n=\{I,X,Y,Z\}^{\otimes n}$ denote the $n$-qubit Pauli basis. Any $n$-qubit operator $A$ can be expressed as a linear combination of Pauli strings,
\begin{equation}
A=\sum_{P\in\mathcal{P}_n}\alpha_P P,
\end{equation}
where \(\alpha_P\) is the coefficient associated with Pauli string $P$. Pauli strings can be efficiently manipulated using the multiplication rules of the single-qubit Pauli operators. For example,
\begin{equation}
    XY=iZ,\qquad YZ=iX,\qquad ZX=iY.
\end{equation}

The single-qubit identities \(X^2=Y^2=Z^2=I\) further imply that the product of any two Pauli strings is another Pauli string up to a phase in $\{1,-1,i,-i\}$. For example,
\begin{equation}
    (X\otimes Z)(Y\otimes X)
=(XY)\otimes(ZX)
=iZ\otimes iY
=-Z\otimes Y.
\end{equation}

These properties allow Pauli strings to be propagated and multiplied without explicitly constructing their matrix representations.
In particular, for a quantum gate $U$, Pauli propagation can be formulated through the conjugation of an operator,
$A\mapsto UAU^\dagger$.
For Clifford gates, conjugation maps every Pauli string to a single Pauli string up to a phase. For example,
\begin{equation}
    HXH=Z,\quad HZH=X,\quad HYH=-Y,\quad
SXS^\dagger=Y,\quad SYS^\dagger=-X,\quad SZS^\dagger=Z.
\end{equation}

Consequently, Clifford gates do not introduce additional Pauli terms during propagation. This property allows a Pauli operator to be represented by a single Pauli string throughout a Clifford circuit. A single Pauli string can branch into multiple terms during propagation. General quantum circuits therefore require a weighted Pauli representation.
By maintaining these weighted Pauli representations sparsely, Pauli propagation can avoid explicit statevector or unitary-matrix construction while preserving the operator-level semantics of the circuit. This combination of Pauli algebra, Clifford conjugation, and sparse term management provides the basis for Pauli-based quantum circuit analysis and equivalence checking.

\section{Methodology}
\label{sec:method}
\begin{figure*}[t]
\centerline{\includegraphics[width=1\linewidth]{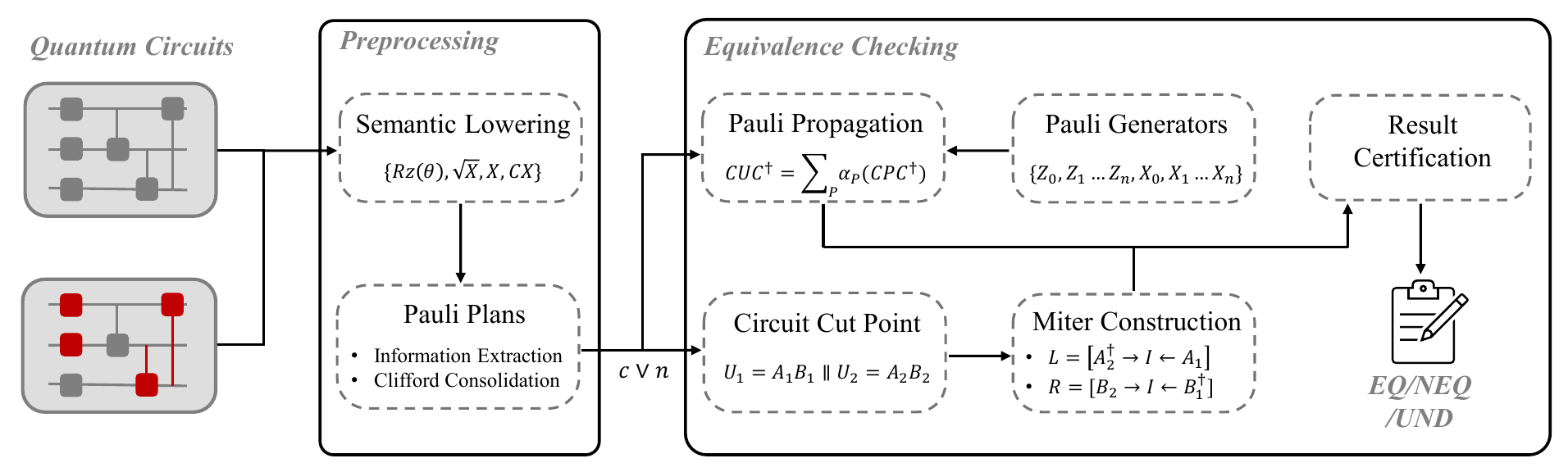}}
\caption{The overall workflow of quantum circuit equivalence checking in our framework.}
\label{fig:workflow}
\vspace{-2mm}
\end{figure*}

\autoref{fig:workflow} illustrates the overall workflow of our framework, which consists of two major modules: (1) the \textit{Preprocessing} module first lowers the input quantum circuit to the basis gate set $\{R_z,\sqrt{X}, X, CX\}$. Under this representation, all non-Clifford effects are captured by $R_z$ rotations, while $\{\sqrt{X},X,CX\}$ are Clifford gates. This normalization simplifies the subsequent Pauli-string expansion by restricting the non-Clifford operations that need to be handled explicitly. The module then extracts structural information from the normalized circuit and consolidates consecutive Clifford gates into Clifford blocks. Non-Clifford $R_z$ operations are kept as individual operations. The resulting representation consists of Clifford blocks and individual non-Clifford operations, thereby exposing the circuit structure for subsequent analysis.
(2) the \textit{Equivalence Checking} module selects different checking strategies based on the number of qubits and the non-Clifford ratio of the input circuits. For circuits with a low non-Clifford ratio or a large number of qubits, we employ the Pauli-generator evolution strategy (Section ~\ref{subsec:generator}). A low non-Clifford ratio ensures that the evolution of each Pauli generator remains efficient, while for large circuits, non-equivalence can be efficiently detected by evolving the finite set of $2n$ Pauli generators. For other cases, we employ the cut-miter strategy (Section ~\ref{subsec:choi}). The circuits are partitioned into multiple segments, and a miter is constructed at each cut point to analyze the weight of the non-identity Pauli components and determine equivalence. To account for numerical errors and the effects of pruning, the results produced by these strategies are further processed by the \textit{Result Certification} module (Section ~\ref{subsec:guarantees}). The certification procedure validates the checking result and ultimately returns the equivalence verdict.

\subsection{Generator-Based Verification in the Heisenberg Picture}
\label{subsec:generator}

Our first backend compares the conjugation actions of the two circuits
on Pauli generators. This representation is insensitive to global
phase and avoids constructing dense unitary matrices or state vectors.
Let $\mathcal{G}_n=
\{Z_1,\ldots,Z_n,X_1,\ldots,X_n\}$
be the canonical generating set of the $n$-qubit Pauli group up to
phase. We use the following standard characterization, which has also
been applied to Clifford-circuit equivalence checking by Thanos et
al.~\cite{thanos2023fast}.

\begin{theorem}[Generator criterion]
\label{lem:generator-completeness}
For arbitrary $n$-qubit unitaries $U_1$ and $U_2$,

\begin{equation}
U_1\equiv U_2
\quad\Longleftrightarrow\quad
U_1 G U_1^\dagger
=
U_2 G U_2^\dagger
\quad
\text{for every }G\in\mathcal{G}_n.
\label{eq:generator-characterization}
\end{equation}

\end{theorem}

\begin{proof}[Proof sketch]
A global phase cancels under conjugation, proving the forward
direction. Conversely, every Pauli operator is, up to phase, a product
of elements of $\mathcal{G}_n$. Agreement on the generators therefore
implies $U_1PU_1^\dagger=U_2PU_2^\dagger$ for every
$P\in\mathcal{P}_n$, where
$\mathcal{P}_n=\{I,X,Y,Z\}^{\otimes n}$. Since every density operator
admits a Pauli expansion
$\rho=\sum_{P\in\mathcal{P}_n}\alpha_PP$, linearity gives

\begin{equation}
    U_1\rho U_1^\dagger
=
\sum_P\alpha_PU_1PU_1^\dagger
=
\sum_P\alpha_PU_2PU_2^\dagger
=
U_2\rho U_2^\dagger.
\end{equation}

For any density operator $\rho$, $U_1$ and $U_2$ satisfy the above equation, demonstrating $U_1 \equiv U_2$.
\end{proof}

The characterization holds for any complete generating set of the Pauli group up to phase. We choose $\mathcal{G}_n$ because it is a minimal set of $2n$ local operators. Consequently, equivalence over all quantum inputs is reduced to $2n$ operator comparisons, and any generator on which Eq.~\eqref{eq:generator-characterization} fails provides an operator-level witness of inequivalence.

\paragraph{Sparse propagation.}
For each $G\in\mathcal{G}_n$, the backend independently propagates $G$ through each of the two input circuits, yielding the two Heisenberg images
\begin{equation}
\mathcal{H}_i(G)
=
U_i G U_i^\dagger
=
\sum_{P\in\mathcal{P}_n}\alpha_{i,G,P}P,
\qquad i\in\{1,2\}.
\label{eq:generator-cloud}
\end{equation}

Each Heisenberg image is maintained as a sparse Pauli representation. Terms producing the same Pauli string are merged by summing their coefficients.

For a Clifford gate $C$, conjugation maps each Pauli string to a single Pauli string up to a sign:
\begin{equation}
    CPC^\dagger = sP',
\qquad
P,P'\in\mathcal{P}_n,\quad
s\in\{-1,+1\}.
\end{equation}

Consequently, Clifford gates preserve the number of Pauli terms and introduce no branching. We exploit this property by grouping maximal consecutive Clifford sequences into single propagation units, which are applied using precomputed conjugation tables.

For each non-Clifford gate, we decompose it into a semantically equivalent composition of $R_z(\theta)$ rotations and Clifford gates. After this transformation, the only non-Clifford operations are $R_z(\theta)$ rotations with angles that are not integer multiples of $\pi/2$. Its local conjugation action is

\begin{equation}
\begin{aligned}
R_z(\theta)IR_z(\theta)^\dagger &= I, &
R_z(\theta)ZR_z(\theta)^\dagger &= Z,\\
R_z(\theta)XR_z(\theta)^\dagger
&=\cos(\theta)X+\sin(\theta)Y, &
R_z(\theta)YR_z(\theta)^\dagger
&=\cos(\theta)Y-\sin(\theta)X.
\end{aligned}
\label{eq:rz-generator-action}
\end{equation}

Hence, an $R_z$ gate branches only active terms containing $X$ or $Y$ on its target qubit; terms containing $I$ or $Z$ remain unchanged. Identical terms created by different branches are subsequently merged. The generator backend therefore propagates Pauli operators through the \emph{conjugation action} of non-Clifford gates, rather than explicitly representing or expanding the gate operators.

\paragraph{Comparison and scheduling.}
After propagating $G$ through both circuits, we quantify the discrepancy between $\mathcal{H}_1(G)$ and $\mathcal{H}_2(G)$ using the $\ell_2$ distance between their coefficient vectors.
\begin{equation} 
r_G = \left\| \boldsymbol{\alpha}_{1,G} - \boldsymbol{\alpha}_{2,G} \right\|_2,
\qquad
\boldsymbol{\alpha}_{i,G} = \left(\alpha_{i,G,P}\right)_{P\in\mathcal{P}_n}, \quad i\in\{1,2\}.
\label{eq:generator-residual} 
\end{equation}
In exact arithmetic, $r_G=0$ if and only if $\mathcal{H}_1(G)=\mathcal{H}_2(G)$. Hence, a nonzero residual for any generator is sufficient to witness non-equivalence, while equivalence is established only when all $2n$ generator images coincide. Propagation attempts process one or more generators through one circuit under term and time limits. The time limit for an attempt is a fixed fraction of the remaining overall time. If an attempt involving multiple generators exceeds either limit, propagation is retried for individual generators. A generator whose propagation remains incomplete after retries is skipped, leaving time to check other generators. Completed comparisons are evaluated using the certification conditions in Section~\ref{subsec:guarantees}. If no verdict is established after these attempts or the overall time limit is reached, the checker returns \textsc{Undecided}.

\subsection{Miter-Based Verification in the Pauli Representation}
\label{subsec:choi}
The generator backend propagates the two circuits independently and compares their Heisenberg images only after propagation. However, it cannot directly exploit cancellations between corresponding regions of the two circuits. Our second verification backend instead compares the circuits through their composite miter $M=U_2^\dagger U_1$.
By the definition of equivalence up to global phase,
\begin{equation}
U_1\equiv U_2
\quad\Longleftrightarrow\quad
\exists,\phi\in\mathbb{R},\
M=e^{i\phi}I.
\label{eq:miter-identity}
\end{equation}
Rather than constructing the $2^n\times2^n$ unitary matrix $M$ explicitly, we characterize the miter condition through its Pauli representation. 
We expand the miter in the Pauli basis
$\mathcal{P}_n=\{I,X,Y,Z\}^{\otimes n}$ as
\begin{equation}
M=
U_2^\dagger U_1=
(\sum_{R\in\mathcal{P}_n}\alpha_R R)(\sum_{Q\in\mathcal{P}_n}\alpha'_Q Q)=
\sum_{R,Q\in\mathcal{P}_n}\alpha_R\alpha'_Q RQ=
\sum_{P\in\mathcal{P}_n}\beta_P P,
\qquad
\beta_P
=
\frac{1}{2^n}\operatorname{Tr}(P^\dagger M).
\label{eq:miter-pauli-expansion}
\end{equation}
Specifically, we represent each quantum gate in $U_1$ and $U_2^\dagger$ as a linear combination of Pauli strings and construct $M$ by performing Pauli-representation multiplication gate by gate. After each multiplication, terms corresponding to the same Pauli string are merged by summing their coefficients, thereby avoiding the explicit construction of its unitary matrix. At this point, the equivalence-checking problem is reduced to checking whether all non-identity Pauli coefficients in the Pauli expansion of \(M\) are zero. However, this approach cannot guarantee sparsity throughout the computation. When represented in terms of Pauli strings, a quantum gate may contain multiple Pauli terms rather than a single term, causing the number of intermediate terms to grow exponentially.

Prior work~\cite{qcec1} uses an alternating miter construction to accelerate equivalence checking of quantum circuits. Following this approach, we incrementally construct the miter from the output boundaries of both circuits, forming a structure of the form $G \rightarrow I \leftarrow G'$. By applying gates from the two circuits from opposite sides, inverse operations corresponding to aligned regions can be brought together and canceled early. This cancellation can keep the intermediate Pauli sum small and reduce the number of terms processed by subsequent updates.

\paragraph{Cut-Miter Pauli Comparison}
However, this alternating construction is most effective when the two circuits remain locally aligned throughout the construction. If the circuits differ substantially around an internal region, the partial miter may grow before the construction reaches equivalent regions on the other side of the difference. To mitigate this limitation, we introduce an internal cut that decomposes each circuit into a prefix and a suffix, and construct two partial products from the two sides of the cut: $U_1=A_1B_1$ and $U_2=A_2B_2$, where \(B_i\) is the prefix before the cut and \(A_i\) is the suffix after it. Rearranging the equivalence condition gives 
\begin{equation}
    U_1=e^{i\phi}U_2 \quad\Longleftrightarrow\quad A_2^\dagger A_1 = e^{i\phi}B_2B_1^\dagger. 
\label{eq:cut-equivalence} 
\end{equation}

Accordingly, the backend constructs the following structure in the form of a miter:
\begin{equation}
    L=[A_2^\dagger \rightarrow I \leftarrow A_1]=\sum_{P\in\mathcal{P}_n}\beta_{L,P} P, \qquad R=[B_2\rightarrow I \leftarrow B_1^\dagger]=\sum_{P\in\mathcal{P}_n}\beta_{R,P} P, 
\end{equation}

using the alternating procedure described above. Cancellation between corresponding suffixes and prefixes can reduce the intermediate Pauli support of \(L\) and \(R\), respectively.
Let $\boldsymbol\ell=(\beta_{L,P})_{P\in\mathcal P_n}$ and $\boldsymbol r=(\beta_{R,P})_{P\in\mathcal P_n}$ be the coefficient vectors of $L$ and $R$. Pauli orthogonality gives $\operatorname{Tr}(P^\dagger Q)=2^n\delta_{P,Q}$. Since both products are unitary, their coefficient vectors have unit norm:
\begin{equation}
    \|\boldsymbol\ell\|_2^2
=\frac{\operatorname{Tr}(L^\dagger L)}{2^n}=1,
\qquad
\|\boldsymbol r\|_2^2
=\frac{\operatorname{Tr}(R^\dagger R)}{2^n}=1.
\end{equation}
For $L=A_2^\dagger A_1$ and $R=B_2B_1^\dagger$, orthogonality and the cyclicity of the trace give
\begin{equation}
    \langle\boldsymbol r,\boldsymbol\ell\rangle
=\sum_{P\in\mathcal P_n}
  \overline{\beta_{R,P}}\beta_{L,P}
=\frac{\operatorname{Tr}(R^\dagger L)}{2^n}
=\frac{\operatorname{Tr}(B_1B_2^\dagger A_2^\dagger A_1)}{2^n}
=\frac{\operatorname{Tr}(M)}{2^n}.
\end{equation}

The inner product is thus independent of the cut position. To account for global phase, we choose
\begin{equation}
    \lambda=
\begin{cases}
\dfrac{\langle\boldsymbol r,\boldsymbol\ell\rangle}
      {|\langle\boldsymbol r,\boldsymbol\ell\rangle|},
& |\langle\boldsymbol r,\boldsymbol\ell\rangle|>0,\\[1.2ex]
1, & \text{otherwise}.
\end{cases}
\end{equation}
This phase minimizes the coefficient distance, which satisfies
\begin{equation}
    d=\|\boldsymbol\ell-\lambda\boldsymbol r\|_2,
\qquad
d^2=2-2|\langle\boldsymbol r,\boldsymbol\ell\rangle|.
\end{equation}
For unitary $M$, $\sum_P|\beta_P|^2=1$ and $\beta_I=\operatorname{Tr}(M)/2^n$. We define the residual as the $\ell_2$ norm of its non-identity coefficients:
\begin{equation}
r_M
=\left(\sum_{P\ne I}|\beta_P|^2\right)^{1/2}
=\sqrt{1-|\langle\boldsymbol r,\boldsymbol\ell\rangle|^2}
=d\sqrt{1-d^2/4}.
\label{eq:cut-residual}
\end{equation}
In exact arithmetic, $r_M=0$ is equivalent to $U_1\equiv U_2$. The cost of constructing $L$ and $R$ can vary with the cut position, so the backend searches over several candidates. Candidate cuts are selected based on the cumulative number of non-Clifford rotations rather than the raw gate count, as non-Clifford rotations are the primary source of Pauli-support growth. The backend first evaluates a balanced cut and then considers asymmetric alternatives. Candidate cuts are initially screened under a small probe budget, and the most promising candidates are subsequently re-evaluated under the full budget. A probe contributes to the final verdict only when the complete coefficient-vector comparison is successfully completed. If a probe exceeds its resource budget before the comparison is completed, it is treated as inconclusive.


\subsection{Complexity and Certified Results}
\label{subsec:guarantees}

\paragraph{Complexity.} 
The resulting operators are propagated in the Pauli basis as sparse Pauli sums. The dominant cost of this propagation comes from the growth of the number of Pauli strings. In particular, a Clifford operation maps each Pauli string to a single Pauli string and therefore does not increase the representation size, whereas a non-Clifford rotation can split one Pauli string into two. Consequently, if a circuit contains $t_i$ non-Clifford rotations, independently propagating the circuit may generate up to $2^{t_i}$ Pauli strings. Let quantum circuits $C_1$ and $C_2$ contain $m$ and $m'$ gates, respectively. As a baseline, independently propagating the two circuits requires $O(m2^{t_1}+m'2^{t_2})$ string operations, where $t_i$ is the total number of non-Clifford rotations in $C_i$. Thus, the baseline cost depends on the global non-Clifford complexity of the circuits.

The cut-miter backend exploits structural similarity between the two circuits to avoid this global expansion. Let $|D|$ be the number of differing rotations and $t_c$ be the maximum number of common non-Clifford rotations encountered by any partial miter. A differing rotation can introduce at most one additional branching factor of two, while common Clifford operations do not increase the number of Pauli strings. Therefore, by the size bound, each partial miter contains at most $2^{|D|+t_c}$ Pauli strings, yielding a total construction cost of $O((m+m')2^{|D|+t_c})$. The key point is that non-Clifford rotations outside the difference region do not contribute to the exponential growth of the partial miter. Hence, when the circuits differ only in a small and localized region, $(|D|+t_c)$ can be substantially smaller than the global non-Clifford count, reducing the number of Pauli strings that must be maintained.

The generator-based backend exploits locality within each circuit rather than similarity between the two circuits. For a generator $G$, let $t_i(G)$ denote the number of non-Clifford rotations in its light cone in circuit $C_i$. Rotations outside this light cone cannot affect the propagation of $G$, so propagating $G$ requires at most $2^{t_i(G)}$ Pauli strings. Thus, the cost for one generator is $O(m2^{t_1(G)}+m'2^{t_2(G)})$, and the total cost over the 2n generators is $O(\sum_G(m2^{t_1(G)}+m'2^{t_2(G)}))$. Although this backend propagates multiple generators, each propagation depends only on the non-Clifford rotations in the corresponding light cone. It can therefore be substantially cheaper than global propagation when these light cones are sufficiently small.

Overall, the two backends exploit different sources of locality: the cut-miter backend exploits similarity between the input circuits, while the generator-based backend exploits locality of individual generators. Neither backend is asymptotically superior for all circuits; their benefits arise when the corresponding local complexity is substantially smaller than the global non-Clifford complexity.

\paragraph{Certified Results.}
The underlying equivalence criterion is exact: two circuits are equivalent if and only if their miter residual is zero. In practice, however, Pauli-term pruning and semantic lowering introduce bounded approximation errors that must be accounted for when making numerical decisions. We therefore use a tolerance $\epsilon$ for decision making while explicitly tracking the error bounds introduced by these approximations. Let $L$ denote the accumulated $\ell_2$-norm of the coefficients removed by pruning from both compared representations, and let $s$ be the combined operator-norm error bound introduced by snapping near-Clifford $R_z$ rotations in both circuits. We use $b=L+s$ for cut-miter comparison and $b=L+2s$ for generator comparison, since snapping changes both $U$ and $U^\dagger$ in $UGU^\dagger$. In exact arithmetic, a residual $q$ computed from the approximate representation satisfies $q-b\leq q_{\mathrm{exact}}\leq q+b$.

For the cut comparison, the corresponding miter residual is $r_M=d\sqrt{1-d^2/4}$.
Therefore, if $r_M+b\leq\epsilon$, the checker can certify the circuits as \textsc{Equivalent}. Conversely, if $r_M-b>\epsilon$, the checker can certify them as \textsc{Not-Equivalent}. Otherwise, the comparison cannot establish a definitive result under the current error bound.

For the generator comparison, let $r_G$ denote the discrepancy between the propagated images of a Pauli generator $G$ in the two circuits. Although the generator backend does not explicitly construct the miter, its certification can be related to the miter residual $r_M$. In exact arithmetic, the discrepancy of each generator satisfies $r_G^{\mathrm{exact}}\leq 2r_M$.
Thus, a sufficiently large discrepancy for any single generator is sufficient to establish non-equivalence. In particular, if
$r_G-b>2\epsilon$,
then $r_G^{\mathrm{exact}}>2\epsilon$, which implies $r_M>\epsilon$. The checker can therefore certify the circuits as \textsc{Not-Equivalent}.
In contrast, certifying equivalence requires a global upper bound on the miter residual, rather than evidence from a single generator. For the complete set of $2n$ Pauli generators, their discrepancies collectively provide such an upper bound, since $2r_M \leq \sqrt{\sum_G\left(r_G^{\mathrm{exact}}\right)^2}$. Taking the approximation error into account, each exact discrepancy is bounded by $r_G+b$. We therefore use $S=\sum_G(r_G+b)^2$ as a conservative upper bound on the squared discrepancies. It follows that $2r_M\leq\sqrt{S}$. Hence, if $\frac{\sqrt{S}}{2}\leq\epsilon$, then $r_M\leq\epsilon$, and the checker can certify the circuits as \textsc{Equivalent}. This certificate requires all 2n generators to complete, because the bound relies on their collective discrepancies.

\subsection{Algorithm}
\label{subsec:algorithm}
This section describes how our framework selects an equivalence-checking strategy according to the structural characteristics of the target quantum circuits. Algorithm~\autoref{alg1} presents the overall workflow of the proposed structure-aware equivalence-checking approach.

\begin{algorithm}[t]
	\renewcommand{\algorithmicrequire}{\textbf{Input:}}
	\renewcommand{\algorithmicensure}{\textbf{Output:}}
	\caption{: Structure-Aware Quantum Circuit Equivalence Checking}
	\label{alg1}
	\begin{footnotesize}
	\begin{algorithmic}[1]
	\REQUIRE Quantum circuits $C_1, C_2$ and residual tolerance $\epsilon$
	\ENSURE \textsc{Equivalent}, \textsc{Not-Equivalent}, or \textsc{Undecided}
	        (\textsc{Timeout} $\Rightarrow$ \textsc{Undecided})

	\IF{$C_1$ and $C_2$ have different numbers of qubits}
		\RETURN \textsc{Not-Equivalent}
	\ENDIF
	\STATE $\Pi_1, \Pi_2 \gets \textsc{PauliPlans}(\textsc{Lowering}(C_1, C_2), \tau_{snap})$
	\STATE  $c \gets \max\{ \Pi_1.\mathrm{non\_clifford},\Pi_2.\mathrm{non\_clifford}\}$
    \STATE $s \gets \Pi_1.\mathrm{snap\_err}+\Pi_2.\mathrm{snap\_err}$

	\IF{$c \ge c_{threshold}$ \AND $n \le n_{threshold}$}
		\FORALL{scheduled cut attempts at fractions $f\in F_{\mathrm{cut}}$}
			\STATE $A_i, B_i \gets \textsc{CutPoint}(f, \Pi_i)$ for $i = 1, 2$
			\STATE $(\boldsymbol{\ell},L_{\boldsymbol{\ell}}), (\boldsymbol{r},L_{\boldsymbol{r}}) \gets
        \textsc{PauliSum}(A_2^\dagger A_1,\tau_{prune}),
        \textsc{PauliSum}(B_2 B_1^\dagger,\tau_{prune})$
            \IF{\textsc{BudgetExceeded}}
            \STATE \textbf{continue}
            \ENDIF
            \STATE $b \gets L_{\boldsymbol{\ell}}+L_{\boldsymbol{r}} + s$
			\STATE $\lambda \gets \langle \boldsymbol{r}, \boldsymbol{\ell} \rangle
			        \,/\, |\langle \boldsymbol{r}, \boldsymbol{\ell} \rangle|$;\ \ $d \gets \|\boldsymbol{\ell} - \lambda \boldsymbol{r}\|_2$;\ \ $r \gets d\sqrt{1 - d^2/4}$
			\RETURN \textsc{Equivalent} if $r + b \le \epsilon$;\ \
			        \textsc{Not-Equivalent} if $r - b > \epsilon$
		\ENDFOR
	\ENDIF

	\STATE $incomplete \gets \FALSE$;\ \ $S \gets 0$
	\FORALL{$G \in \{Z_1, \ldots, Z_n, X_1, \ldots, X_n\}$}
		\STATE $\boldsymbol{\alpha}_1, \boldsymbol{\alpha}_2,L_1, L_2 \gets
		        \textsc{SparsePropagation}(G, \Pi_1, \Pi_2, \tau_{prune})$
        \IF{\textsc{BudgetExceeded}}
    \STATE $incomplete \gets \TRUE$
    \STATE \textbf{continue}
            \ENDIF
        \STATE $b \gets L_1+L_2+2s $
		\IF{$\|\boldsymbol{\alpha}_1 - \boldsymbol{\alpha}_2\|_2 - b > 2\epsilon$}
			\RETURN \textsc{Not-Equivalent}
		\ENDIF
		\STATE $S \gets S + (\|\boldsymbol{\alpha}_1 - \boldsymbol{\alpha}_2\|_2 + b)^2$
	\ENDFOR
	\RETURN \textsc{Undecided} if $incomplete$ or $\sqrt{S}/2 > \epsilon$ else \textsc{Equivalent}
	\end{algorithmic}
	\end{footnotesize}
\end{algorithm}

Given two quantum circuits $C_1$ and $C_2$ and a residual tolerance $\epsilon$, the checker first verifies that the two circuits act on the same number of qubits (lines 1-3). It then performs semantic lowering and constructs the corresponding Pauli propagation plans $\Pi_1$ and $\Pi_2$ (line 4). The maximum proportion of non-Clifford operations in the two plans is used as the structural indicator $c$ (line 5), while the accumulated operator-norm error $s$ introduced by snapping is also obtained (line 6).


When the circuits contain a sufficiently large proportion of non-Clifford operations and the number of qubits is within the predefined threshold (line 7), the checker first invokes the cut-based backend. It evaluates a set of candidate cut positions, with the balanced cut considered first and asymmetric cuts subsequently examined (line 8). For each candidate cut, the circuits are decomposed into prefix and suffix parts (line 9), and the two corresponding Pauli-sum representations are constructed. During Pauli-sum construction, we prune coefficients whose magnitudes are below $\tau_{prune}$ and return the accumulated error introduced by the pruning process (line 10). If the construction exceeds the string budget, the current cut is skipped, and the next candidate is considered (line 11-13). Otherwise, the accumulated approximation error is bounded by $b=L+s$ (line 14), where $L$ denotes the pruning error and $s$ denotes the semantic-lowering error. The coefficient vectors $\boldsymbol{\ell}$ and $\boldsymbol{r}$ are aligned up to a global phase $\lambda$, and their cut distance $d$ is computed. This distance is converted into the corresponding miter residual $r=d\sqrt{1-d^2/4}$ (line 15). Accounting for the approximation bound $b$, the checker immediately certifies the circuits as \textsc{Equivalent} when $r+b\leq\epsilon$, or as \textsc{Not-Equivalent} when $r-b>\epsilon$ (line 16). If neither condition holds, the current cut is inconclusive, and the checker proceeds to the next candidate.

If the cut-based backend is not applicable or cannot establish a definitive result, the checker falls back to the generator-based backend. It propagates each of the $2n$ Pauli generators through both circuits and obtains the corresponding coefficient vectors $\boldsymbol{\alpha}_1$, $\boldsymbol{\alpha}_2$ and pruning error $L$ (line 21). If propagation for a generator exceeds the string budget, the checker marks the analysis as incomplete and continues with the remaining generators (line 22-25). For every successfully propagated generator, the discrepancy $r_G=|\boldsymbol{\alpha}_1-\boldsymbol{\alpha}_2|_2$ is computed. If $r_G-b>2\epsilon$, the generator provides a sufficient witness for non-equivalence (line 27), and the checker immediately returns \textsc{Not-Equivalent} (line 28). Otherwise, the conservative upper bound $(r_G+b)^2$ is accumulated into $S$ (line 30). After all generators have been processed, the checker can certify equivalence only when every generator has been successfully analyzed and the resulting global bound satisfies $\sqrt{S}/2\leq\epsilon$. If some generators were incomplete or this condition is not satisfied, the checker returns \textsc{Undecided} (line 32). 
Thus, the algorithm exploits an asymmetric certification principle: a single sufficiently large generator discrepancy is sufficient to certify non-equivalence, whereas equivalence requires complete coverage of all generators and a sufficiently small aggregate discrepancy. 

\textit{Implementation.}
We use $\epsilon=10^{-10}$ as the default residual tolerance. Pauli coefficients with magnitude at most $\tau_{\mathrm{prune}}=10^{-14}$ are removed after merging duplicate terms. This threshold eliminates floating-point cancellation residue while remaining well below $\epsilon$. 
During Pauli-plan construction, an $R_z$ rotation within $\tau_{\mathrm{snap}}=10^{-12}$ of a Clifford angle is snapped to that angle. The actual perturbation introduced by these snaps is accumulated as $s$. All verdicts use the certified error band $b$. The cut backend is enabled when the larger plan contains at least $c_{\mathrm{threshold}}=16$ non-Clifford rotations and the circuits have at most $n_{\mathrm{threshold}}=32$ qubits. The former is an empirical routing threshold that avoids cut-search overhead on rotation-sparse circuits. The latter follows from the current implementation, which stores the $X$ and $Z$ masks of a Pauli string in one 64-bit word. We try the cut fractions $F_{\mathrm{cut}}=(0.50,0.35,0.65,0.25,0.75,0)$, where $0$ represents the input-end one-sided construction, $0.5$ is the balanced cut, and the remaining fractions cover asymmetric alignments. If the cut backend cannot certify a verdict within its resource budget, the algorithm falls back to generator propagation.

\section{Experimental Evaluation}
\label{sec:experiment}
Our framework is implemented in Python using Qiskit~2.5.1~\cite{qiskit}. We address the following research questions:

\definecolor{bleudefrance}{rgb}{0.19, 0.55, 0.91}

\begin{itemize}
    \item \noindent\textbf{RQ1: Effectiveness and Efficiency.}
    How effectively and efficiently does our framework check quantum circuit equivalence compared with existing approaches?

    \item \noindent\textbf{RQ2: Component Effectiveness.}
    How does each component contribute to the overall performance of our approach?

    \item \noindent\textbf{RQ3: Pruning Sensitivity.}
    How sensitive is our approach to the pruning threshold $\tau_{\mathrm{prune}}$?
\end{itemize}

In \textbf{RQ1}, we compare our framework with QCEC, Quokka\#, and QuPRS on two benchmark suites. We evaluate their correctness, decision coverage, and runtime under the same timeout limit. In \textbf{RQ2}, we conduct an ablation study of individual components of our framework. We evaluate configurations without cut-miter checking, with only the input-end cut, with the input-end cut evaluated first, without per-attempt time limits, and without Pauli-term pruning. We compare their decision coverage, runtime, and memory consumption with those of the complete configuration. In \textbf{RQ3}, we evaluate our framework with $\tau_{\mathrm{prune}}\in\{10^{-14},10^{-13},10^{-12}\}$. We study how the threshold affects decision coverage and runtime.

\noindent\textit{Baselines.}
MQT QCEC~\cite{qcectool} combines decision-diagram-based construction, alternating checking, simulation, and ZX-calculus rewriting. Quokka\#~\cite{quokka} reduces Pauli-generator equivalence obligations to weighted model counting.
The Quokka\# baseline represents the WMC-based equivalence-checking implementation provided by the Quokka\# framework, which builds on the line of work introduced by ECMC~\cite{ecmc}.
QuPRS~\cite{quprs} combines path-sum reduction with a weighted-model-counting fallback. We use QCEC~3.8.0, Quokka\#~2.7.5 on Union and the artifact-provided version~2.7.3 on Artifact, and QuPRS~0.13.0 in hybrid mode. QCEC uses its default configuration on Union, while its ZX checker is disabled on Artifact to match the original artifact's experimental settings~\cite{quokka}.

\noindent\textit{Benchmarks.}
Our primary benchmark \textbf{Union} contains 350 circuit pairs derived from 70 base circuits. The base circuits cover 14 MQT Bench families, including GHZ, Deutsch--Jozsa, graph-state, W-state, QFT, QPE, QAOA, VQE, amplitude-estimation, quantum-walk, Grover, and QNN circuits, together with random Clifford+$T$ circuits. Circuit widths range from 4 to 512 qubits. For each base circuit $C$, we remove final measurements and obtain $C_{\mathrm{opt}}$ using Qiskit at optimization level 2, with the gate set $\{\mathrm{id},R_z,\sqrt{X},X,\mathrm{CX}\}$ and seed 42. We compare $C$ with $C_{\mathrm{opt}}$ and four variants generated separately from $C_{\mathrm{opt}}$ by deleting one gate (\texttt{del}), swapping the control and target of one CX gate (\texttt{flip}), or increasing one $R_z$ angle by $10^{-4}$ (\texttt{shift4}) or $10^{-7}$ (\texttt{shift7}). This construction yields five pairs (1 equivalent circuit pair classified as EQ and 4 non-equivalent circuit pairs classified as NEQ) per base circuit, with mutation positions selected using a fixed seed for each circuit family and width. Before freezing the resulting QASM files for reuse across experiments, we normalize the final qubit order and translate both sides at optimization level 0. Our framework, QCEC, and QuPRS use the gate set above. Quokka\#~2.7.5 uses $\{R_x,R_z,H,\mathrm{CX}\}$ because it does not support \texttt{sx} and incorrectly encodes $R_y$ rotations. The \textbf{Artifact} benchmark contains 88 circuit pairs distributed with the Quokka\# artifact. These include the original optimization, gate-missing, and rotation-perturbation pairs, covering both algorithmic circuits and reversible arithmetic benchmarks.

\noindent\textit{Experimental setup.}
All experiments were conducted on a machine running NixOS 25.11, equipped with a 14-core Intel Core i9-10940X CPU at 3.30\,GHz, with 28 hardware threads and 204\,GiB of RAM. We allow 300 seconds per circuit pair for all tools. Our framework uses an equivalence tolerance of $10^{-10}$ and a default pruning threshold of $10^{-14}$. Generator propagation attempts are limited to 10\% of the remaining time. The corresponding RQ2 variant removes the per-attempt limit while retaining the overall timeout.

\subsection{RQ1: Effectiveness and Efficiency}

\begin{table}[htbp]
    \centering
    \caption{Comparison with existing equivalence checkers by reference label.}
    \label{tab:rq1-overall}
    \small
    \setlength{\tabcolsep}{3pt}
    \begin{tabular*}{\linewidth}{@{\extracolsep{\fill}}lcccrcccr@{}}
        \toprule
        & \multicolumn{4}{c}{Union (346 instances)} & \multicolumn{4}{c}{Artifact (88 instances)} \\
        \cmidrule(lr){2-5}\cmidrule(l){6-9}
        Tool & Correct & Incorrect & Inconclusive & Time (s) & Correct & Incorrect & Inconclusive & Time (s) \\
        \midrule
        & \multicolumn{4}{c}{EQ (66 instances)} & \multicolumn{4}{c}{EQ (7 instances)} \\
        \cmidrule(lr){2-5}\cmidrule(l){6-9}
        \textbf{Ours} & 64 & 0 & 2 & 572.6 & 7 & 0 & 0 & 24.0 \\
        QCEC & 64 & 0 & 2 & 611.2 & 7 & 0 & 0 & 6.0 \\
        Quokka\# & 49 & 0 & 17 & 6\,064.8 & 6 & 0 & 1 & 457.8 \\
        QuPRS & 52 & 0 & 14 & 4\,362.1 & 5 & 0 & 2 & 756.3 \\
        \midrule
        & \multicolumn{4}{c}{NEQ (280 instances)} & \multicolumn{4}{c}{NEQ (81 instances)} \\
        \cmidrule(lr){2-5}\cmidrule(l){6-9}
        \textbf{Ours} & 279 & 0 & 1 & 1\,987.6 & 69 & 0 & 12 & 5\,183.1 \\
        QCEC & 174 & 42 & 64 & 8\,468.1 & 36 & 21 & 24 & 7\,550.9 \\
        Quokka\# & 188 & 49 & 43 & 15\,478.2 & 55 & 4 & 22 & 7\,366.6 \\
        QuPRS & 108 & 108 & 64 & 19\,405.1 & 10 & 33 & 38 & 11\,224.9 \\
        \bottomrule
    \end{tabular*}
    \par\smallskip
    \begin{minipage}{\linewidth}
        \footnotesize
        Groups and correctness are based on reference labels. Inconclusive denotes no EQ or NEQ result within 300 seconds. Time is summed within each group.
    \end{minipage}
\end{table}

\begin{table}[htbp]
    \centering
    \caption{Representative instances illustrating complementary strengths.}
    \label{tab:rq1-cases}
    \small
    \setlength{\tabcolsep}{3pt}
    \begin{tabular*}{\linewidth}{@{\extracolsep{\fill}}lcccc@{}}
        \toprule
        Instance & Ours & QCEC & Quokka\# & QuPRS \\
        \midrule
        DJ-64, \texttt{shift7} (U) & \textbf{NEQ (0.036s)} & EQ$^{\dagger}$ (0.194s) & EQ$^{\dagger}$ (1.441s) & EQ$^{\dagger}$ (1.068s) \\
        VQE-2L-10, \texttt{shift7} (U) & \textbf{NEQ (0.067s)} & T/O & T/O & T/O \\
        QWalk-8, \texttt{shift4} (A) & \textbf{NEQ (24.118s)} & T/O & T/O & T/O \\
        QPEexact-16, \texttt{flip} (U) & NEQ (180.930s) & T/O & \textbf{NEQ (0.698s)} & NEQ (0.791s) \\
        QNN-16, \texttt{eq} (U) & EQ (29.315s) & \textbf{EQ (0.048s)} & EQ (0.461s) & EQ (3.686s) \\
        \bottomrule
    \end{tabular*}
    \par\smallskip
    \begin{minipage}{\linewidth}
        \footnotesize
        U and A denote Union and Artifact. T/O denotes timeout; $\dagger$ marks an incorrect result.
    \end{minipage}
    \vspace{-2mm}
\end{table}

\paragraph{Overall comparison.}
Our method returns reference-consistent results for 343 of 346 scored Union instances and 76 of 88 Artifact instances, with no incorrect results and the lowest aggregate runtime on both suites (Table~\ref{tab:rq1-overall}). Our method and QCEC correctly classify 64 of 66 EQ instances on Union and all seven on Artifact. All additional correct results over QCEC are on NEQ instances. On Union, the 105 additional results resolve 42 misclassifications and 63 inconclusive cases. On Artifact EQ instances, QCEC takes 6.0\,s compared with our 24.0\,s. The four excluded Union instances are the \texttt{eq} variants of QFT-16, QPEexact-32, QPEinexact-32, and QPEinexact-64. Their reference labels assume that O2 optimization preserves equivalence, while Qiskit applies numerical tolerances even with \texttt{approximation\_degree=1.0}~\cite{eval_qiskit16864}. The resulting differences exceed our circuit tolerance, and both our method and Quokka\# return NEQ on these pairs.

\paragraph{Structural suitability.}
Table~\ref{tab:rq1-cases} illustrates how circuit structure affects cost. DJ-64 contains Clifford gates apart from the perturbed rotation, so generator conjugation produces at most two Pauli terms per generator despite its width. Partial-product comparison can also keep intermediate Pauli representations small when common circuit structure cancels. On VQE-2L-10 with \texttt{shift7}, the balanced cut establishes NEQ in 0.067\,s, with a recorded peak of 4\,112 Pauli terms, while all three baselines time out. On QWalk-8 with \texttt{shift4}, generator comparison establishes NEQ in 24.118\,s using a residual of $1.768\times10^{-5}$.

The baselines can be substantially faster on other circuit structures. QCEC uses decision-diagram methods and simulation. Quokka\# encodes Pauli propagation as weighted Boolean constraints, and QuPRS simplifies path-sums before invoking model counting. A large Pauli expansion may still admit efficient reasoning in these representations. On QPEexact-16 with \texttt{flip}, our method requires generator retries and takes 180.930\,s, whereas Quokka\# and QuPRS both finish below one second. QCEC checks QNN-16 \texttt{eq} in 0.048\,s, compared with our 29.315\,s. Across the 238 Union instances, both QCEC and our method classify correctly; the median runtime ratio $T_{\rm QCEC}/T_{\rm ours}$ is 1.22.

\paragraph{Numerical sensitivity.}
On the 70 Union \texttt{shift7} instances, our method detects all 70, compared with 21 for QCEC, four for Quokka\#, and zero for QuPRS. The DJ example illustrates how the comparison quantity affects detection. An affected generator has residual $2|\sin(\delta/2)|\approx|\delta|$, whereas Quokka\#'s projected coefficient departs from one by $1-\cos\delta\approx\delta^2/2$. For $\delta=10^{-7}$, the latter is approximately $5\times10^{-15}$, below its configured numerical tolerance of $10^{-12}$. 
The original ECMC evaluation also considered phase-shift perturbations of $10^{-4}$ and $10^{-7}$, and also reports incorrect results for $10^{-7}$ cases. The evaluated QuPRS implementation also rounds the base-two logarithm of the WMC magnitude to three decimal places before its tolerance test~\cite{eval_quprs_source}, reducing the resolution of that comparison. Our method compares the computed residual with the circuit tolerance while accounting for errors from pruning and semantic lowering. 
These results indicate that, for the small phase perturbations considered in our evaluation, residual-based comparison provides a higher-resolution discriminative signal, while explicit error bounds allow our framework to retain this numerical sensitivity without overlooking the approximation errors introduced during computation.

\begin{tcolorbox}[size=title,rightrule=1mm,leftrule=1mm,toprule=0mm,bottomrule=0mm,arc=0pt,colback=gray!5,colframe=bleudefrance!75!black,breakable]
\textbf{Answer to RQ1: }
Our framework correctly decides more instances than the evaluated state-of-the-art equivalence checkers on Union and Artifact, with the lowest aggregate runtime. No incorrect results are observed among the scored instances.
\end{tcolorbox}

\subsection{RQ2: Component Effectiveness}
We compare the full method with five variants that separately remove partial-product comparison, restrict the candidate cuts, change their order, disable pruning, or remove per-attempt time limits for generator propagation (Table~\ref{tab:rq2-ablation}). All other settings remain fixed.

\begin{table}[t]
    \centering
    \caption{Ablation study on 350 Union instances.}
    \label{tab:rq2-ablation}
    \small
    \setlength{\tabcolsep}{3pt}
    \begin{tabular*}{\linewidth}{@{\extracolsep{\fill}}lrrrr@{}}
        \toprule
        Configuration & Decided & Time (s) & Rel. time & Peak RSS (GiB) \\
        \midrule
        Full method & \textbf{347} & \textbf{2\,643} & 1.00$\times$ & \textbf{12.25} \\
        - Without cut stage & 344 & 4\,739 & 1.79$\times$ & 43.14 \\
        - Without per-attempt time limits & 345 & 3\,232 & 1.22$\times$ & 32.03 \\
        - Without pruning & 318 & 14\,884 & 5.63$\times$ & 109.21 \\
        Input-end first & \textbf{347} & 2\,999 & 1.13$\times$ & 12.41 \\
        Input-end miter & \textbf{347} & 3\,070 & 1.16$\times$ & 12.41 \\
        \bottomrule
    \end{tabular*}
    \par\smallskip
    \begin{minipage}{\linewidth}
        \footnotesize
        All variants retain generator verification. Input-end first retains all candidate cuts and evaluates $f=0$ first, while input-end miter uses only $f=0$. Decided counts EQ or NEQ results returned within 300\,s. Time is summed within each group. Peak RSS is the process high-water mark, including execution beyond the deadline.
    \end{minipage}
    \vspace{-2mm}
\end{table}

Removing the cut stage leaves generator verification alone, loses three EQ results, and increases aggregate runtime by a factor of 1.79. Recorded peak RSS also rises from 12.25 to 43.14\,GiB, a factor of 3.52. The input-end miter decides all three pairs and records a peak of 12.41\,GiB. Joint construction at the input end therefore already provides the additional decisions and most of the observed memory reduction. Comparing circuit products permits cancellation between the circuits and can establish a result before separate generator expansions are required.

The full method reduces aggregate runtime by 11.9\% and 13.9\% relative to input-end first and input-end miter, respectively, with the same 347 decisions. Internal cuts improve efficiency on this suite without increasing coverage beyond the input-end miter. Input-end first uses the same candidate cuts as the full method, so their order also affects the benefit obtained within the shared cut-stage budget. This benefit is concentrated: the Grover-8 \texttt{gm}, \texttt{flip}, and \texttt{shift7} instances account for 108.0\% of the net time reduction over input-end first. Trying an expensive boundary first can consume the cut-stage budget before an inexpensive comparison is completed.

Disabling pruning reduces the number of decided instances from 347 to 318, increases aggregate runtime by a factor of 5.63, and raises recorded peak RSS from 12.25 to 109.21\,GiB. Retained terms must be processed at later propagation steps, where non-Clifford rotations can expand them further. Pruning reduces the work and storage required by subsequent updates as well as the size of the current sum.

Removing per-attempt time limits reduces the number of decided instances from 347 to 345 and increases aggregate runtime by 22.3\%. Both additional decisions are NEQ, for which one generator comparison can suffice. Limiting expensive propagation attempts leaves time for other generators and can improve coverage without increasing the overall timeout.

To further understand how the two verification backends contribute to the overall decision coverage, we analyze the 350 \textbf{Union} circuit pairs according to the backend that produces the final decision. This analysis distinguishes pairs decided directly by the cut-miter stage from those decided by generator verification, as well as pairs for which the cut stage is inconclusive.

\begin{table}[t]
    \centering
    \caption{Deciding stages of the full configuration on Union benchmark.}
    \label{tab:rq2-stages}
    \small
    \setlength{\tabcolsep}{5pt}
    \begin{tabular*}{0.6\linewidth}{@{\extracolsep{\fill}}lr@{}}
        \toprule
        Stage or outcome & Instances \\
        \midrule
        Cut-miter checking & 200 \\
        Generator verification & 147 \\
        \textsc{Undecided} & 3 \\
        \midrule
        Total & 350 \\
        \bottomrule
    \end{tabular*}
    \vspace{-2mm}
\end{table}

Table~\ref{tab:rq2-stages} reports 200 decisions made directly by cut-miter, 147 decisions made by generator verification, and three undecided pairs. Among the 147 generator-verification decisions, 115 are obtained without any cut attempt, while the remaining 32 follow inconclusive cut attempts. Thus, among the 235 circuit pairs subjected to the cut stage, 200 are decided directly by cut-miter, while the remaining 35 proceed to generator verification; 32 of these are subsequently decided, and three remain undecided. Overall, the two backends are complementary: cut-miter resolves 200 pairs directly, while generator verification resolves an additional 147 pairs. This distribution indicates that generator verification is not merely a fallback when cut-miter cannot reach a decision, but independently contributes a substantial portion of the final decisions.

Further analysis by circuit family shows that the 235 pairs subjected to the cut stage include 100 QAOA, QFT, and VQE pairs, all of which are decided by cut-miter, but only 15 of the 35 QPE pairs are decided at this stage. Generator verification decides 17 of the remaining 20 QPE pairs, leaving three undecided. Thus, the current cut fractions are more effective for the tested QAOA, QFT, and VQE pairs than for QPE. Selecting cuts using circuit structure and estimates of intermediate Pauli-term growth may allow more pairs to be decided before generator propagation.


\begin{tcolorbox}[size=title,rightrule=1mm, leftrule=1mm, toprule=0mm, bottomrule=0mm, arc=0pt,colback=gray!5,colframe=bleudefrance!75!black,breakable]
{ \textbf{Answer to RQ2: } The ablation results show that the components of our framework contribute differently to coverage and efficiency. On the Union benchmark, the cut-miter and generator verification backends are complementary, resolving 200 and 147 pairs, respectively. Pruning has the largest impact: disabling it reduces coverage from 347 to 318 and increases runtime by 5.63$\times$. Per-attempt time limits and cut ordering provide additional benefits, while internal cuts primarily improve runtime.
}
\end{tcolorbox}

\subsection{RQ3: Pruning Sensitivity}
\paragraph{Pruning-threshold sensitivity.}
Our framework relies on numerical thresholds in its Pauli propagation and decision procedure. In particular, the pruning threshold $\tau_{\mathrm{prune}}$ determines which Pauli coefficients are discarded after merging duplicate terms. A larger threshold discards more coefficients and can therefore increase the accumulated pruning error bound, which may in turn prevent a comparison from establishing EQ or NEQ. At the same time, a larger threshold can reduce the number of Pauli terms retained for subsequent propagation and thus lower the computational cost. We therefore evaluate the sensitivity of our framework to $\tau_{\mathrm{prune}}$ by comparing three thresholds, $10^{-12}$, $10^{-13}$, and $10^{-14}$, on both Union and Artifact. We keep the circuit-equivalence tolerance fixed and use the full configuration in all runs. We compare both decision outcomes and aggregate runtime to determine whether reducing the pruning threshold affects decision stability or provides additional coverage at a measurable computational cost.

\paragraph{Outcome stability.}
Across the three threshold settings in Table~\ref{tab:rq3-pruning}, 437 of 438 instances retain the same output, with no EQ--NEQ switches. The only difference occurs for one instance that is inconclusive at $10^{-12}$ but is classified as EQ at both smaller thresholds. Thus, within the tested range, changing the pruning threshold has little effect on the final decision outcomes, and no instance switches directly between EQ and NEQ.

\paragraph{Computational trade-off.}
Reducing the pruning threshold from $10^{-13}$ to $10^{-14}$ decreases aggregate runtime by 0.6\% on Union and by less than 0.1\% on Artifact in these runs, with no change in decision coverage. Combined with the much larger performance degradation observed when pruning is completely disabled in RQ2, these results indicate that pruning itself is important for controlling computational cost on these benchmarks, whereas the exact choice of pruning threshold has relatively little impact within the tested range. The default threshold of $10^{-14}$ therefore provides stable decision outcomes with similar computational cost to $10^{-13}$.

Overall, these results indicate that our framework is relatively insensitive to the pruning threshold within the tested range. Lowering the threshold does not materially change the decision outcomes or improve coverage, while the presence or absence of pruning has a much larger effect on computational performance. This observation supports the use of a small pruning threshold to maintain stable decisions while retaining the computational benefits of pruning.

\begin{table}[t]
    \centering
    \caption{Sensitivity to the pruning threshold.}
    \label{tab:rq3-pruning}
    \small
    \setlength{\tabcolsep}{3pt}
    \begin{tabular*}{0.75\linewidth}{@{\extracolsep{\fill}}lrrrr@{}}
        \toprule
        & \multicolumn{2}{c}{Union} & \multicolumn{2}{c}{Artifact} \\
        \cmidrule(lr){2-3}\cmidrule(l){4-5}
        $\varepsilon_{\rm prune}$ & Decided & Time (s) & Decided & Time (s) \\
        \midrule
        $10^{-12}$ & 346/350 & 2\,627 & 76/88 & 5\,200 \\
        $10^{-13}$ & 347/350 & 2\,660 & 76/88 & 5\,208 \\
        $10^{-14}$ (default) & 347/350 & 2\,643 & 76/88 & 5\,207 \\
        \bottomrule
    \end{tabular*}
    \vspace{-2mm}
\end{table}

\begin{tcolorbox}[size=title,rightrule=1mm, leftrule=1mm, toprule=0mm, bottomrule=0mm, arc=0pt,colback=gray!5,colframe=bleudefrance!75!black,breakable]
{ \textbf{Answer to RQ3: } 
Our framework achieves similar decision coverage and aggregate runtime on Union and Artifact across pruning thresholds from $10^{-14}$ to $10^{-12}$. Reducing the threshold from $10^{-13}$ to $10^{-14}$ preserves the same coverage with similar runtime.
}
\end{tcolorbox}

\section{Threats to Validity}
\label{sec:validity}
\paragraph{Internal validity.}
Implementation errors and finite-precision arithmetic may affect the results. We mitigate these risks through unit tests and numerical matrix comparisons with Qiskit on small circuits. Our framework also accounts for errors introduced by pruning and rotation-angle adjustments when making tolerance-based decisions.

\paragraph{External validity.}
Our evaluation includes both generated circuit pairs and the original Quokka\# artifact. These benchmarks cover different circuit families, sizes, and structures, but may not represent all practical quantum programs. In particular, our framework is primarily designed to exploit structural regularities commonly arising in compiler optimizations and circuit transformations. Its effectiveness may therefore be more limited for arbitrary circuit pairs with little structural correspondence. Mutation positions are fixed, and performance may vary with the mutation location or available resources.

\paragraph{Construct validity.}
Our framework checks full unitary equivalence up to global phase and does not support mid-circuit measurements, classical control, or non-trivial initial layouts. Numerical approximations during transpilation may produce inaccurate reference labels. We check four such discrepancies using Quokka\# and exclude them from all metrics in the RQ1 comparison. Since exact ground truth is unavailable for some large circuits, further labeling errors may remain.

\section{Related Work}
\label{sec:related}

In this section, we review existing methods for quantum circuit equivalence checking and verification beyond unitary circuits. We then discuss Pauli-based simulation and verification methods and their relation to the operator propagation used in our framework.

\subsection{Quantum Circuit Equivalence Checking}

Quantum circuit equivalence checking has been studied using decision diagrams, the ZX-calculus, path sums, and weighted model counting. Burgholzer and Wille~\cite{ec3,qcec1} proposed a decision-diagram-based approach to quantum circuit equivalence checking, in which quantum circuit matrices are represented using decision diagrams. They further introduced a miter-based construction that interleaves operations from the two circuits, allowing common substructures to cancel early and thereby reducing the number of decision-diagram nodes. Peham et al.~\cite{qcec2} proposed a ZX-calculus-based approach that represents quantum circuits as connected graphs and determines equivalence through graph simplification. They also pointed out that this approach does not guarantee completeness. These two approaches have since been integrated into MQT QCEC~\cite{qcectool,mqt} as complementary equivalence-checking techniques. Mei et al.~\cite{ecmc} used the Pauli-generator characterization of unitary equivalence to reduce circuit comparison to weighted model counting (WMC). Huang et al.~\cite{quprs} combined path-sum reductions with WMC, applying model counting to the residual path-sum when simplification does not decide equivalence. Yu et al.~\cite{pbec} used local projections to check equivalence of shallow circuits, with complexity linear in qubit count for fixed depth but exponential in depth for one-dimensional nearest-neighbor circuits. Our framework extends the alternating miter construction used in decision-diagram-based checking with internal cuts, comparing two partial products directly instead of expanding the complete miter. This allows cancellation within both circuit prefixes and suffixes, which can reduce intermediate Pauli-term growth when corresponding regions are aligned. In contrast to WMC-based checking, it computes these products and generator images by direct sparse propagation.

\subsection{Equivalence and Verification Beyond Unitary Circuits}

Equivalence checking becomes more involved for quantum programs containing mid-circuit measurements, resets, discarded qubits, noise, or measurement-dependent classical control. Such programs are generally described by quantum channels or ensembles of branch operators rather than a single unitary transformation. Burgholzer and Wille~\cite{burgholzer2022nonunitary} proposed transformations for handling non-unitary operations while retaining compatibility with existing unitary equivalence-checking techniques. Hong et al.~\cite{hong2022dynamic} formalized the semantics of dynamic quantum circuits using ensembles of linear operators and developed tensor-decision-diagram-based algorithms for checking both measurement-output and quantum-output equivalence. For noisy circuits, Hong et al.~\cite{hong2021noisy} defined approximate equivalence using the Jamiolkowski fidelity between quantum channels and evaluated it through tensor-network contraction. More recently, Ricciardi et al.~\cite{ricciardi2025hybrid} combined deferred measurement with separation and projection techniques to extend unitary equivalence checking to a broader class of hybrid circuits, including circuits with discarded subsystems.

Beyond circuit-level equivalence, formal verification methods reason about general behavioral properties of quantum programs. Ying et al.~\cite{ying2013verification} modeled quantum programs using superoperators and quantum Markov chains and developed methods for verifying observable properties and termination. Barthe et al.~\cite{barthe2020relational} introduced a relational program logic based on quantum couplings, enabling properties relating two program executions to be formally proved. These studies handle non-unitary behavior or general program properties, while our framework focuses on the computational cost of full unitary equivalence checking up to global phase.

\subsection{Pauli-Domain Circuit Verification}

Pauli and stabilizer representations provide an efficient foundation for reasoning about quantum circuits with substantial Clifford structure. Gottesman~\cite{gottesman1998heisenberg} introduced a Heisenberg-picture description in which quantum computation is characterized through the evolution of Pauli operators. Aaronson and Gottesman~\cite{aaronson2004improved} subsequently developed efficient tableau-based algorithms for stabilizer circuits. These algorithms exploit the fact that conjugation by a Clifford gate maps each Pauli string to a single Pauli string. Gidney developed Stim~\cite{stim}, an efficient stabilizer simulator based on tableau representations and Pauli-frame propagation.

Extending stabilizer techniques beyond Clifford circuits generally requires representing the effect of non-Clifford operations using weighted combinations. Bravyi et al.~\cite{bravyi2019stabilizer} represented quantum states as low-rank sums of stabilizer states, enabling the simulation of circuits containing a moderate number of non-Clifford gates. Pashayan et al.~\cite{pashayan2022fast} developed complementary algorithms for estimating output probabilities of circuits containing Clifford gates and arbitrary-angle Pauli-$Z$ rotations. In the operator domain, Rall et al.~\cite{rall2019pauli} formulated circuit simulation through Pauli propagation, where states or observables are evolved directly in the Pauli basis. More recent approaches propagate Heisenberg observables as sparse weighted sums of Pauli strings and use coefficient- or weight-based truncation to control term growth. Begu\v{s}i\'c and Chan~\cite{begusic2025sparse} applied this approach to real-time operator evolution, while Rudolph et al.~\cite{rudolph2025pauli} systematized the underlying algorithms and provided the PauliPropagation.jl framework.

For equivalence checking, Thanos et al.~\cite{thanos2023fast} compare the conjugation of the canonical Pauli generators $X_i$ and $Z_i$ using Stim as the stabilizer-simulation backend. Their approach is efficient but restricted to Clifford circuits. Osama et al.~\cite{osama2025parallel} subsequently parallelized generator-based Clifford equivalence checking on GPUs. Our framework propagates weighted Pauli sums through non-Clifford rotations, while the Clifford generator checkers propagate single Pauli strings. The sparse simulators above compute states or observables, whereas our framework compares a complete set of generator images to establish equivalence over all inputs.

\section{Conclusion}
\label{sec:conclusion}

We presented a structure-aware equivalence checker for unitary quantum circuits based on sparse Pauli propagation. By combining generator propagation with Cut-Miter comparison, our framework adapts its verification strategy to different circuit structures while using bounded resources and explicitly accounting for numerical and approximation errors.
Our evaluation on two benchmark suites with circuits of up to 512 qubits shows that our framework provides correct, reference-consistent decisions for 96.3\% of 434 scored circuit pairs without incorrect verdicts, compared with at most 68.7\% for the evaluated baselines. It also achieves the lowest aggregate runtime, reducing total runtime by 48.5\%--76.0\%. The ablation results further demonstrate that cut-based comparison and Pauli-term pruning are important for maintaining decision coverage and computational efficiency, while bounded propagation attempts improve the use of the available verification budget.

Overall, our results show that exploiting complementary forms of sparsity provides a practical way to make Pauli-based equivalence checking more scalable and robust across diverse quantum circuit structures. Our framework therefore offers a useful verification oracle for quantum software testing and compiler validation. These results also suggest several directions for future work, including structure-adaptive cut-point selection, support for circuits with mid-circuit measurements and classical control, and hybrid verification approaches that combine our framework with SAT-based reasoning and weighted model counting.



\bibliographystyle{ACM-Reference-Format}
\bibliography{AAA_Paper/ref}

\end{document}